\documentclass{article}
\usepackage{spconf,amsmath,graphicx}
\usepackage{amssymb}
\usepackage{pifont}

\usepackage{multirow}
\usepackage{xcolor, tabu, cite}
\newcommand{\xmark}{\ding{55}}%
\title{Time-domain speech extraction with spatial information \\ and multi speaker conditioning mechanism}
\name{Jisi Zhang$^1$, C\u{a}t\u{a}lin Zoril\u{a}$^2$, Rama Doddipatla$^2$ and Jon Barker$^1$
\thanks{\copyright 2021 IEEE. Accepted for ICASSP 2021.}}
\address{$^1$University of Sheffield, Department of Computer Science, Sheffield, UK \\
        $^2$Toshiba Cambridge Research Laboratory, Cambridge, UK}
        
\begin{document}
\ninept
\maketitle
\begin{abstract}
In this paper, we present a novel multi-channel speech extraction system to simultaneously extract multiple clean individual sources from a mixture in noisy and reverberant environments.
The proposed method is built on an improved multi-channel time-domain speech separation network which employs speaker embeddings to identify and extract multiple targets without label permutation ambiguity. To efficiently inform the speaker information to the extraction model, we propose a new speaker conditioning mechanism by designing an additional speaker branch for receiving external speaker embeddings.
Experiments on 2-channel WHAMR! data show that the proposed system improves by 9\% relative the source separation performance over a strong multi-channel baseline, and it increases the speech recognition accuracy by more than 16\% relative over the same baseline.
\end{abstract}
\section{Introduction}
\label{sec:intro}

Speech separation aims to segregate individual speakers from a mixture signal, and it can be used in many applications, such as speaker diarization, speaker verification or multi-talker speech recognition.
Deep learning has allowed an unprecedented separation accuracy compared with the traditional signal processing based methods, however, there are still challenges to address.
For instance, in blind source separation, the order of the output speakers is arbitrary and unknown in advance, which forms a speaker label permutation problem during training.
Clustering based methods~\cite{hershey2016deep} or, more recently, Permutation Invariant Training (PIT) technique~\cite{kolbaek2017multitalker} have been proposed to alleviate this issue.
Although the PIT forces the frames belonging to the same speaker to be aligned with the same output stream, frames inside one utterance can still flip between different sources, leading to a poor separation performance.
Alternatively, the initial PIT-based separation model can be further trained with a fixed label training strategy~\cite{yang2020interrupted}, or a long term dependency can be imposed to the output streams by adding an additional speaker identity loss~\cite{drude2018deep,nachmani2020voice}.
Another issue in blind source separation is that the speaker order of the separated signals during inference is also unknown, and needs to be identified by a speaker recognition system.

An alternative solution to the label permutation problem is to perform target speaker extraction~\cite{vzmolikova2019speakerbeam,Delcroix_2020,Ge2020SpExAC}.
In this case, the separation model is biased with information about the identity of the target speaker to extract from the mixture.
Typically, a speech extraction system consists of two networks, one to generate speaker embeddings, and another one to perform speech extraction.
The speaker embedding network outputs a speaker representation from an enrollment signal uttered by the target.
The speaker embedding network can be either jointly trained with the speech extraction model to minimise the enhancement loss or trained on a different task, i.e., a speaker recognition task, to access larger speaker variations~\cite{Wang_2019}.
The target speaker embedding is usually inserted into the middle-stage features of the extraction network by using multiplication~\cite{Delcroix_2020} or concatenation operations~\cite{Ge2020SpExAC,ji2020speaker},
however, the shared middle-features in the extraction model may not be optimal for both tasks of speaker conditioning and speech reconstruction.

Most of the existing speech extraction models enhance only one target speaker each time and ignore speech from other speakers.
When multiple speakers are of interest, the extraction model has to be applied several times, which is inconvenient and requires more computational resources.
Therefore, a system capable of simultaneously extracting multiple speakers from a mixture is of practical importance.
Recently, a speaker-conditional chain model (SCCM) has been proposed that firstly infers speaker identities, then uses the corresponding speaker embeddings to extract all sources~\cite{shi2020speaker}.
However, SCCM is still trained with the PIT criterion, and the output order of separated signals is arbitrary. 
Lastly, when multiple microphones are available, the spatial information has been shown to improve the performance of both separation and extraction~\cite{Zhang2020end,Delcroix_2020} systems in clean and reveberant environments.
So far, the spatial information has not been tested with a multi-speaker extraction system, nor it has been evaluated in noisy and reverberant environments.

In this paper, we reformulate our previous multi-channel speech separation design in~\cite{Zhang2020end} as a multi-talker speech extraction system.
The proposed system uses embeddings from all speakers in the mixture to simultaneously extract all sources, and does not require PIT to solve the label permutation problem.
There are three main contributions in this work.
Firstly, we improve our previous multi-channel system in~\cite{Zhang2020end} by swapping the Temporal fully-Convolutional Network (TCN) blocks with U-Convolutional blocks, which yielded promising results for a recent single-channel speech separation model~\cite{tzinis2020sudo}.
Secondly, the previous modified system is reformulated to perform multi-speaker extraction, and, lastly, a novel speaker conditioning mechanism is proposed that exploits the speaker embeddings more effectively.
The evaluation is performed with multi-channel noisy and reverberant 2-speaker mixtures.
We show that combining the updated multi-channel structure and the proposed speaker conditioning mechanism leads to a significant improvement in terms of both the separation metric and speech recognition accuracy.

The rest of paper is organised as follows.
In section~\ref{sec:multi_extr}, we introduce the proposed multi-channel speech extraction approach.
Section~\ref{sec:experiment} presents implementation details and the experiment setup.
Results and analysis are presented in Section~\ref{sec:result}.
Finally, the paper is concluded in Section~\ref{sec:conclusion}.

\section{Multi-channel end-to-end extraction}
\label{sec:multi_extr}

\begin{figure}[htp]
    \centering
    \includegraphics[width=0.48\textwidth]{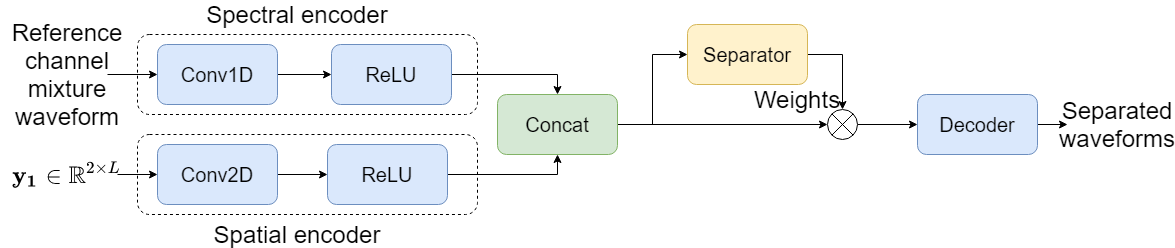}
    \caption{Updated multi-channel model structure}
    \label{fig:update_multi}
\end{figure}

Recently, neural network based multi-channel speech separation approaches have achieved state-of-the-art performance by directly processing time-domain speech signals~\cite{Zhang2020end,gu2020enhancing}.
These systems incorporate a spectral encoder, a spatial encoder, a separator, and a decoder.
In~\cite{Zhang2020end}, spatial features are input to the separator only.
In this work, we simplify the previous framework by combining the spatial and spectral features as depicted in Figure~\ref{fig:update_multi}.
We found the proposed approach is beneficial for the speech extraction task.
The spectral encoder and spatial encoder independently generate $N$-dimensional single-channel representations and $S$-dimensional multi-channel representations, respectively.
The spectral encoder is a 1-D convolutional layer, and the spatial encoder is a 2-D convolutional layer.
The encoded single-channel spectral features and two-channel spatial features 
are concatenated together to form multi-channel representations with a dimension of $(N+S)$, which are accessed by both the separation module and the decoder.
The separator will estimate linear weights for combining the multi-channel representations to generate separated representations for each source.
Finally, the decoder (1-D convolutional layer) reconstructs the estimated signals by inverting the separated representations back to time-domain signals.

\begin{figure}[htp]
    \centering
    \includegraphics[width=0.45\textwidth]{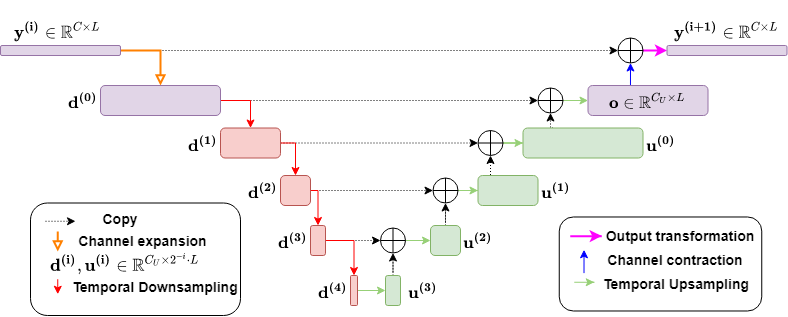}
    \caption{U-Conv block structure}
    \label{fig:u_conv}
\end{figure}

\begin{figure}[htp]
    \centering
    \includegraphics[width=0.45\textwidth]{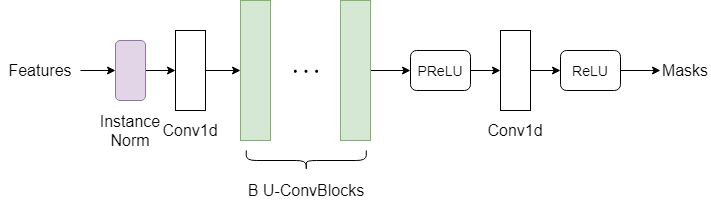}
    \caption{Improved separator with U-Conv blocks}
    \label{fig:u_conv_sep}
\end{figure}

Compared with our previous work~\cite{Zhang2020end}, we also upgrade the separator by replacing the original TCN~\cite{lea2016temporal} blocks with U-Convolutional blocks (U-ConvBlock), which have proven to be more effective in modelling sequential signals in the single-channel speech separation task~\cite{tzinis2020sudo}.
Furthermore, a system built on U-ConvBlock requires fewer parameters and floating point operations compared with the systems built on TCN or recurrent neural network architectures~\cite{luo2020dual}.
The U-ConvBlock (Figure~\ref{fig:u_conv}) extracts information from multiple resolutions using $Q$ successive temporal downsampling and $Q$ upsampling operations similar to a U-Net structure~\cite{ronneberger2015u}.
The channel dimension of the input to each U-ConvBlock is expanded from $C$ to $C_U$ before downsampling, and is contracted to the original dimension after upsampling.
The updated separation module is shown in Figure~\ref{fig:u_conv_sep} and consists of a instance normalisation layer, a bottleneck layer, $B$ stacked U-ConvBlocks and a 1-D convolutional layer with a non-linear activation function.
We choose to use an instance normalisation layer~\cite{ulyanov2016instance} rather than global layer normalisation for the first layer-normalisation, 
as the latter would normalise over the channel dimension which is inappropriate given the heterogeneous nature of the concatenated features.

\subsection{Proposed speech extraction structure}

Building on the modified system described above, in this section we introduce a novel multi-channel speech extraction system which simultaneously tracks multiple sources in the mixture.
In general, the system uses embeddings from multiple speakers as input, which are used to condition single-source outputs with a consistent speaker order.
Common strategies for supplying speaker information to the extraction model are to modulate the speaker features on middle-level features inside the separation model~\cite{vzmolikova2019speakerbeam,zeghidour2020wavesplit} or concatenate the speaker features with the mixture speech representations~\cite{Ge2020SpExAC}.
However, it is not trivial to find a single optimal layer at which to insert the speaker features.
For instance, the shared middle-features in the extraction model may not be optimal for both speaker conditioning and speech reconstruction.

\begin{figure}[b]
    \centering
    \includegraphics[width=0.48\textwidth]{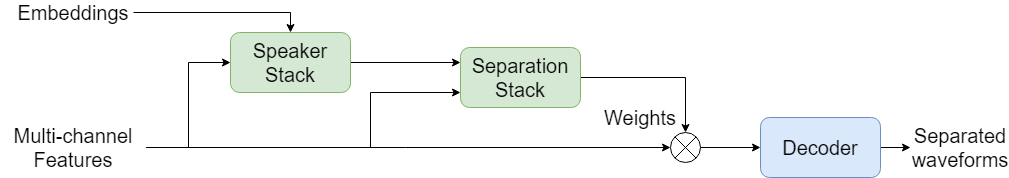}
    \caption{Proposed multi-channel speech extractor with dedicated speaker stack}
    \label{fig:multi_channel_split_extr}
\end{figure}
\begin{figure}[hb!]
    \centering
    \includegraphics[width=0.48\textwidth]{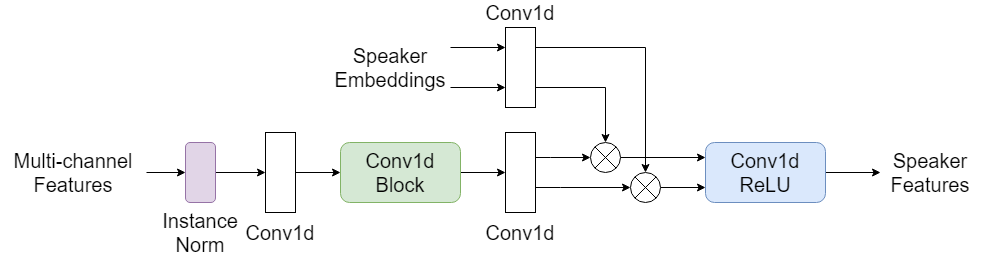}
    \caption{Internal structure of proposed speaker stack}
    \label{fig:spkr_stack}
\end{figure}

To address this issue, we propose a new `speaker stack' for processing the input speaker representations to coordinate with the main separation stack, as shown in Figure~\ref{fig:multi_channel_split_extr}.
The speaker stack takes the encoded multi-channel features and generates two high-level sequential features, which are suitable to receive speaker information from externally computed speaker embeddings.
The output of the speaker branch containing speaker information is encouraged to learn similar characteristics as the original multi-channel features and can be concatenated together as input to the separation stack.
Note that the encoder is shared for both the speaker stack and the separation stack.
The speaker stack, illustrated in Figure~\ref{fig:spkr_stack}, first employs an instance normalisation, a bottleneck 1-D CNN and a single TCN block to receive multi-channel features.
Then, the output of the TCN block will be factorised by an adaptation layer into multiple features for modulation with multiple speaker embeddings, which are transformed with a $1 \times 1$ convolutional layer to the same feature dimension.
The modulated signals from each speaker embedding are concatenated together and processed with a 1-D convolutional layer and a ReLU non-linear activation function to form $E$-dimensional speaker information features, which have the same time length as the multi-channel features.

The speaker stack and the separation stack are jointly trained to directly optimise the scale-invariant signal-to-noise ratio (SI-SNR) metric~\cite{le2019sdr},
\begin{equation}
    \begin{split}
    & \text{SI\text{-}SNR} = 10\mathrm{log}_{10} \frac{||\text{s}_{target}||^2}{||\text{e}_{noise}||^2} \\
    & \text{s}_{target} = \frac{\big \langle \hat{s}, s \big \rangle s}{||s||^2},  \quad
    \text{e}_{noise} = \hat{s} - s_{target} 
    \end{split}
    \label{equ:si_snr}
\end{equation}
where $\hat{s}$ and $s$ denote the estimated and clean source, respectively, and $||s||^2=\big \langle s, s \big \rangle$ denotes the signal power.
In contrast with PIT, we condition the decoded signals on the speaker representations and keep the output speaker order consistent with the order of input speaker embeddings.

\section{Experiment Setup}
\label{sec:experiment}
\subsection{Data simulation}
The evaluation is performed on the WHAMR! dataset~\cite{Maciejewski_2020}, which consists of simulated noisy and reverberant 2-speaker mixtures.
WHAMR! is based on Wall Street Journal (WSJ) data, mixed with noise recorded in various urban environments~\cite{Wichern_2019}, and artificial room impulse responses generated by using pyroomacoustics~\cite{scheibler2018pyroomacoustics} to approximate domestic and classroom environments.
There are 20k sentences from 101 speakers for training, and 3k sentences from 18 speakers for testing.
The speakers in the test set do not appear during training of the speaker recognition model nor they appear during training of the speaker extraction system.
All data are binaural (2-channels) and have 8 kHz sampling rate.

\subsection{Speech extraction network}
The multi-channel separation network in~\cite{Zhang2020end} trained with PIT has been set as the baseline for comparison.
The hyper-parameters of the baseline model are the same as those for the best model in the original paper, chosen as follows, $N=256$, $S=36$, $R=3$, $X=7$, $L=20$, and the batch size $M=3$.
For the U-ConvBlock based separation module, the hyper-parameters are set as SuDoRM-RF 1.0x in~\cite{tzinis2020sudo} namely, $L=21$, $B=16$, $Q=4$, $C=256$, $C_U=512$, and the training batch size $M=4$.
Each utterance is split into multiple segments with a fixed length of 4 seconds.
The dimension of speaker features, $E$, in the speaker stack is set to 128.
The ADAM optimizer~\cite{kingma2014adam} is used for training with a learning rate of $1e-3$, which will be halved if the loss of validation set is not reduced in 3 consecutive epochs.
All models are trained with 100 epochs.
The input for all the models is the reveberant mixture with noise and the targets are the clean individual sources.

\subsection{Speaker recognition network}
We retrained the time-domain speaker recognition model SincNet \cite{Ravanelli_2018} for speaker embedding generation. Employing the same configuration as in the original paper,
SincNet is trained on the clean training set of WSJ0 (101 speakers), using speech segments of 200~ms with 10~ms overlap.
The output of the last hidden layer of final SincNet model represents one frame-level speaker embedding for each 200 ms segment, and an utterance-level embedding is derived by averaging all the frame predictions.

Randomly selecting a single enrollment utterance for generating the speaker embedding leads to poor extraction performance.
Therefore, to increase the robustness, we follow an averaging strategy to obtain one global embedding for each speaker~\cite{Li2019Target}.
Specifically, each global speaker embedding is obtained by averaging several embeddings generated from multiple randomly selected utterances belonging to the same speaker.
During training, one global speaker embedding is generated by averaging all the utterance-level embeddings from the training utterances belonging to the corresponding speaker.
During evaluation, 3 utterances are randomly selected for each speaker, and the utterance-level embeddings from the selected utterances are averaged to form one global embedding.
Experiments showed that increasing the number of utterances beyond 3 does not improve performance.

\subsection{Acoustic model}
To evaluate the speech recognition performance, two acoustic models have been trained using the WSJ corpus. One model (AM1) was trained on roughly 80~hrs of clean WSJ-SI284 data plus the WHAMR! single-speaker noisy reverberant speech, and the other one (AM2) was trained on the data used for AM1 plus the separated signals from the WHAMR! mixture in the training set processed by the proposed model.
The audio data is downsampled to 8 kHz to match the sampling rate of data used for separation experiments.
The acoustic model topology is a 12-layered Factorised TDNN~\cite{povey2018semi}, where each layer has 1024 units.
The input to the acoustic model is 40-dimensional MFCCs and a 100-dimensional i-Vector.
A 3-gram language model is used during recognition.
The acoustic model is implemented with the Kaldi speech recognition toolkit~\cite{povey2011kaldi}.
With our set-up, the ASR results obtained with AM1 on the standard clean WSJ Dev93 and Eval92 are 7.2\% and 5.0\% WER, respectively.
\vspace{-5pt}
\section{Results and Analysis}
\label{sec:result}
\vspace{-5pt}
\subsection{Improved Multi-channel separation network}
Table~\ref{tab:whamr_sep_perf} reports the separation performance for the improved multi-channel separation network with various configurations.
The first observation is that the dimension of the spatial features does not have to be fixed to a small value (typically 36) as mentioned in the previous work.
The results show that when the dimension increases, more useful spatial information is extracted and the model benefits more from the multi-channel signals.
Replacing the TCN blocks with the stacked U-ConvBlocks provides a larger receptive field due to successive downsampling operations, and the latter model yields 0.5 dB SI-SNR improvement.
The configuration depicted in the last row of Table 1 is used for the rest of the experiments.

\begin{table}[htp]
\centering
\caption{Speech separation performance of improved multi-channel structure on WHAMR! test set}
\label{tab:whamr_sep_perf}
\begin{tabular}{lcc}
\hline
\textbf{Model}                         & \textbf{S}    & \textbf{SI-SNRi}  \\ \hline
Multi-TasNet (TCN)                     & 36            & 12.1             \\ 
Multi-TasNet (TCN)                     & 64            & 12.2             \\ 
Multi-TasNet (TCN)                     & 128           & 12.4             \\ \hline
Multi-TasNet (U-Conv)                  & 128           & 12.9             \\ \hline
\end{tabular}
\end{table}

\subsection{Results of speech extraction system}
Three subsets of experiments with different speaker information conditioning strategies are performed.
The first experiment uses the multiplication strategy applied in SpeakerBeam~\cite{Delcroix_2020}, which modulates the speaker embedding on the middle-stage representations in the separation module, denoted as Multiply.
The second experiment repeats and concatenates the speaker embeddings with the spectral and spatial representations before being fed into the separation module, denoted as Concat.
Lastly, the third experiment uses the proposed conditioning mechanism, denoted as Split.
\vspace{-12pt}
\begin{table}[htp]
\centering
\caption{Speech extraction performance with improved multi-channel structure on the WHAMR! test set}
\label{tab:whamr_extr_perf}
\begin{tabular}{lccc}
\hline
\textbf{Model}                           & \textbf{PIT} & \textbf{SI-SNRi} \\ \hline
Separation (Improved)                    & \checkmark           & 12.9     \\
Extraction (Concat)                      & \xmark               & 12.8     \\
Extraction (Multiply)                    & \xmark               & 12.9      \\ \hline
Extraction (Split)                       & \xmark               & 13.3     \\
Extraction (Split)                       & \checkmark           & 13.4     \\ \hline
\end{tabular}
\end{table}

The results in Table~\ref{tab:whamr_extr_perf} show that the extraction model cannot directly benefit from the speaker information through the multiplication or concatenation strategies.
The reason for failure of direct multiplication is presumed to be that the shared middle-stage features are not optimal for both tasks of speaker conditioning and speech reconstruction.
As for the concatenation, the multi-channel features and the speaker embedding are completely different signals and cannot be suitably processed by the convolutional layer, which assume time and frequency homogeneity.
Conversely, the separation model with the proposed mechanism can benefit from the speaker information and outperforms the blind source separation system and other conditioning strategies.
The proposed method uses a separated speaker branch to generate high-level features for speaker conditioning tasks to alleviate the shared feature problem.
And the sequential speaker features from the speaker branch can have a similar signal characteristic to the multi-channel features, which is a suitable input to the convolutional layers.

It should be noted that the proposed speech extraction system can be evaluated without accessing reference clean speech to find the right permutation.
When the system is evaluated with the PIT criterion to find the oracle permutation, there is only a small difference between the two results.
This demonstrates that our system can successfully identify and track multiple speakers in noisy and reverberant acoustic conditions.
\vspace{-12pt}
\begin{table}[htp]
\centering
\caption{Results on different and same gender mixtures}
\label{tab:whamr_gender_analysis}
\begin{tabular}{lcccc}
\hline
\multirow{2}{*}{\textbf{Model}} & \multirow{2}{*}{\#nchs} & \multirow{2}{*}{\textbf{PIT}} & \multicolumn{2}{c}{\textbf{SI-SNRi}} \\ \cline{4-5} 
                                &       &                               & Diff.             & Same             \\ \hline
SuDo-RMRF~\cite{tzinis2020sudo}                       & 1     & \checkmark                    & 10.6              & 9.1              \\
Multi-TasNet (TCN)              & 2     & \checkmark                    & 12.4              & 12.4             \\
Multi-TasNet (U-Conv)           & 2     & \checkmark                    & 12.9              & 12.9              \\
Extraction (Split)              & 2     & \xmark                        & 13.5              & 13.1              \\
Extraction (Split)              & 2     & \checkmark                    & 13.5              & 13.3             \\ \hline
\end{tabular}
\vspace{-5pt}
\end{table}

Table~\ref{tab:whamr_gender_analysis} reports the performance of various systems with different and same gender WHAMR! mixture speech.
For blind source separation, a single-channel system can achieve better separation performance with different gender mixtures than same gender mixtures.
With the spatial information, a multi-channel system improves performance in both conditions and reduces the gap between the two mixture conditions.
With the additional speaker information, the performance in the different gender condition is further boosted.
It can be also noticed that the same gender mixtures are more challenging, and more future work is needed to find better speaker representations in this case.

Table~\ref{tab:whamr_all_perf} compares the proposed approach with other competing systems evaluated on WHAMR!.
The proposed speaker conditioning mechanism provides consistent separation performance gain in both single and multi-channel scenarios.
With the additional information from multiple microphones and speaker enrollment, our system achieves the best performance.

\vspace{-12pt}
\begin{table}[htp]
\centering
%\caption{Speech separation and extraction performance with improved multi-channel structure on the speaker-unaware condition}

% CZ: What is "the speaker-unaware condition"?
\caption{Comparative results of single and multi-channel speech separation/extraction on WHAMR! data}

\label{tab:whamr_all_perf}
\begin{tabular}{lcccc}
\hline
\textbf{Model}                          & \#nchs & \textbf{Building Unit} & \textbf{PIT} & \textbf{SI-SNRi} \\ \hline
Conv-TasNet~\cite{luo2019conv}          & 1      & TCN                    & \checkmark           & 9.3               \\
SuDo-RMRF~\cite{tzinis2020sudo}         & 1      & U-Conv                 & \checkmark           & 9.9              \\
Wavesplit~\cite{zeghidour2020wavesplit} & 1      & TCN                    & \checkmark           & 12.0              \\
Nachmanis's~\cite{nachmani2020voice}    & 1      & RNN                    & \checkmark           & 12.2              \\ 
Multi-TasNet~\cite{Zhang2020end}        & 2      & TCN                    & \checkmark           & 12.1              \\ \hline
Extraction (Split)                      & 1      & U-Conv                 & \xmark               & 11.1              \\
Extraction (Split)                      & 1      & U-Conv                 & \checkmark           & 11.1              \\
Extraction (Split)                      & 2      & U-Conv                 & \xmark               & 13.3              \\
Extraction (Split)                      & 2      & U-Conv                 & \checkmark           & 13.4              \\ \hline
\end{tabular}
\vspace{-12pt}
\end{table}

\vspace{-12pt}
\begin{table}[htp]
\centering
\caption{Speech recognition results}
\label{tab:whamr_wer}
\begin{tabu}{lccc}
\hline
\multirow{2}{*}{\textbf{System}} & \multirow{2}{*}{\#nchs} & \multicolumn{2}{c}{\textbf{WER(\%)}} \\ \cline{3-4} 
                                 &                        & AM1               & AM2              \\ \hline
Mixture                          & -                      & 79.1              & 77.0             \\
Multi-TasNet~\cite{Zhang2020end} & 2             & 37.7              & -                \\
Extraction (Split)               & 2            & \bf{31.6}              & \bf{20.9}             \\ \hline
\rowfont{\color{gray}}
Noisy Oracle                     & -             & 19.8              & 20.0             \\ \hline
\end{tabu}
\end{table}

Table~\ref{tab:whamr_wer} reports the ASR results.
The proposed speech extraction model yields a significant WER reduction over the noisy reverberant mixture and outperforms the strong multi-channel separation baseline.
The extraction system can introduce distortions to the separated signals (causing a mismatch problem between training and testing of the acoustic model), therefore, by decoding the data with AM2, the WER is further reduced by 34\% relatively, which is close to the result obtained with oracle single-speaker noisy reverberant speech (last row in Table~\ref{tab:whamr_wer}).

In future work, we plan to exploit other speaker recognition models for embedding generation, and to train these models with larger and more challenging datasets, such as VoxCeleb~\cite{Chung2018}.
Moreover, we will investigate joint training of the speaker embedding and the proposed speech extraction networks, which is expected to benefit both tasks~\cite{ji2020speaker}.
\vspace{-5pt}
\section{Conclusions}
\label{sec:conclusion}
\vspace{-5pt}
In this paper, we have presented a multi-channel speech extraction system with a novel speaker conditioning mechanism.
By introducing an additional speaker branch for receiving external speaker features, this mechanism solves the problems caused by feature sharing from contradicting tasks and difference between multiple inputs, providing a more effective way to use the speaker information to improve separation performance.
Informed by multiple speaker embeddings, the proposed system is able to simultaneously output corresponding sources from a noisy and reverberant mixture, without a label permutation ambiguity.
Experiments on WHAMR! simulated 2-speaker mixtures have shown that the proposed multi speaker extraction approach outperforms a strong blind speech separation baseline based on PIT.
\vfill\pagebreak

% -------------------------------------------------------------------------
\bibliographystyle{IEEEtran}
\bibliography{strings,refs}

\end{document}